\documentclass[twocolumn,final]{article}
\usepackage{a4wide}
\usepackage{showkeys}
\usepackage{amsmath, amssymb, epsf, euscript}
\usepackage{nrj}
\usepackage{graphics,graphicx}

\title{Gradient Based Routing in Wireless Sensor Networks: a Mixed Strategy}
\author{Olivier Powell, Aubin Jarry, Pierre Leone, Jos\'e Rolim
\thanks{\{powell$|$jarry$|$leone$|$rolim\}@cui.unige.ch}}
\begin{document}
\maketitle
\begin{abstract}
We show how recent theoretical advances from
\cite{singh,nrj:balance:jose,nrj:balance:pierre,nrj:balance:olivier}
for data-propagation in Wireless Sensor Networks (WSNs) can be combined to improve
gradient-based routing (GBR) in Wireless Sensor Networks.
We propose a \emph{mixed-strategy} of \emph{direct} transmission and
multi-hop propagation of data which improves the lifespan of WSNs by
reaching better energy-load-balancing amongst sensor nodes.
\cite{GBR}.
\begin{center}
\end{center}
\end{abstract}
\section{Introduction}
An important objective in the design and implementation of efficient algorithms for Wireless Sensor Networks (WSN)
is to minimize the energy consumption, in order to maximize the
lifespan of the network (\cite{WSNSurvey}).
We are interested in  the problem of \emph{data propagation} in
WSNs (\cite{routingSurvey1,routingSurvey2}), which is the fact of transferring information
scattered amongst the nodes of the WSN towards the base station, also called the
sink. In this context, energy is consumed mainly during
the radio transmission phase \cite{}. The energy-cost of transmissions typically grows as a power (greater than $2$) 
of the transmission range (\cite{routingSurvey1,routingSurvey2}). 
Because of this polynomial growth, multi-hop data propagation from node to node towards 
the sink is usually preferred to long-range and expensive single-hop transmission from distant nodes 
directly to the sink, unless the distance from nodes to the base station is small (c.f. \cite{heinz}).
This introduces the necessity to \emph{route} messages from node to node, towards the sink.
One of the possible approaches is gradient based routing (GBR), which was first introduced in \cite{GBR}
building upon ideas from \cite{intana1}. This line of research is still very active,
c.f. for example \cite{gbrMobility,gbrCost}.
One limitation of the multi-hop data propagation approach is that it seems inevitable that
bottle-neck nodes (or bottle-neck regions) in the routing graph tend to be overused and run 
out of energy before others (\cite{singh,nrj:balance:jose,nrj:balance:pierre,nrj:balance:olivier}). 
Because bottle-neck nodes are also central
gateways in the routing graph of the network, energy burn-out of these nodes also leads 
to the premature die-out of the whole network, therefore,
\emph{spreading techniques} 
are considered (already in \cite{GBR})
to force data to diffuse more evenly in the network.
However well spread the data, it still seems inevitable that regions close to the sink are overused,
since they have to route data from the whole network (\emph{a contrario} to nodes at the periphery of the network).
Recently, mixed strategies combining both \emph{multi-hop} data-propagation towards the sink and \emph{direct-transmission}
have shown capable of extending the network lifespan for GBR-like data
propagation, 
c.f. \cite{singh,nrj:balance:jose,nrj:balance:pierre,nrj:balance:olivier}.
We show how these ideas can be combined with those from GBR-like routing  
to get a \emph{mixed-strategy} for data propagation and how this improves the network lifespan.
\section{A Mixed Strategy Combined with Gradient-based Routing}
In \cite{GBR}, a gradient-based routing technique is
proposed. The idea is that each node gets to know its {\bf\emph{height}}, 
which is equal to the hop distance of the node to the base-station (BS).
The gradient from a node to another is then defined as their difference in height, and messages are routed along the
stiffest slope. This simple approach leads to an overuse of a few routes, so they introduce
\emph{network spreading techniques}, in order to put to contribution the rest of the network. For example,
the \emph{stochastic scheme} introduces some randomization in the choice of routes (this idea has also 
been  used in cost field approach, e.g. in \cite{gbrCost}). Two other spreading techniques were also proposed:
an \emph{energy-based} adjustment to the height, and a \emph{stream-based} adjustment. The techniques employed are
general enough so that they are also applicable to other routing protocols.
Nevertheless, however well spread the traffic, the fact that it is routed multi-hop towards the sink
yields an overuse of nodes close to the sink. This problem was acknowledged in \cite{singh}, and further investigated
in the series of papers \cite{nrj:balance:jose,nrj:balance:pierre,nrj:balance:olivier}. In particular, in
\cite{nrj:balance:olivier}, an \emph{offline} algorithm is proposed to compute a \emph{mixed-strategy} combining multi-hop
data-propagation and parsimonious use of direct transmission to the sink, (and it is shown that under some assumptions, 
the mixed-strategy computed by the algorithm optimizes the network lifespan). The algorithm in \cite{nrj:balance:olivier}
works offline and considers an \emph{abstraction} of the GBR
routing scheme. This paper inspires itself from this algorithm to propose a modified version of GBR and show
experimentally the advantages of this approach.\\
{\bf Our algorithm:} In order to adapt GBR to a mixed-strategy, we define a static potential function for each node $n$,
which is equal to the sum of the squared \emph{heights}: $pot_s(n) =
\sum_1^n height^2(n)$. We use an energy-based spreading technique,
by defining the potential of a node as $pot(n) = pot_s(n) + \epsilon(n)$, where $\epsilon(n)$ is a function measuring 
the energy spent so-far by the node. Finally, the mixed strategy is obtained by applying the following scheme: 
\emph{when a node needs to route a message, it sends it to the neighbour with the lowest potential. 
However, if this lowest neighbour has a (strictly) higher potential
than its own, it does not pass the message to a neighbour
but send it \emph{directly} to the the base station}. 
This direct transmission is a long-range transmission, and it has an
energy cost\footnote{The cost is defined using the square of the height, 
but the appropriate power would be larger in noisy a environment. 
This holds for the static potential function $pot_s$ too.} of $height^2(n)$ 
(whereas transmission to a neighbour is assumed to cost $1$ energy unit).
\section{Simulations}
\subsection{Details of the Implementation}
\label{implementation}
For the simulation of our distributed algorithm, we have used the following implementation scheme.
First, we choose a dispersion radius $R$ and the number of sensor nodes $N$ we want to disperse in the
disc $D_R$ centered at the origin. Then we disperse uniformly the $N$ sensors in $D_r$ and add one or more
base stations in $D_R$. 
\begin{enumerate}
\item
In an initialisation phase, each sensor gets to know its height by flooding from the bases stations 
(taking the min-height amongst its base stations). We also assume each
sensor has access to the height and energy spent so-far of its neighbours. 
\item
The run phase is divided in discrete rounds. 
In each round, every sensor which has at least one message in its message queue
sends it to its lowest-potential neighbour (or directly to the sink).
In our implementation, transmission is simultaneous, without collisions, 
and at most one message is sent by each node during each round
(other messages are queued for later transmission). 
In each round, we also \emph{generate an event}: a sensor is picked
at random and its message queue is incremented by $1$. 
\end{enumerate}
In order to compare our algorithm to other possible approaches, and to gain deeper understanding of the dynamics of the network, we run
simultaneous data-propagation protocols on the same WSN and in the same environment 
(i.e. with the same generated events).
In particular, we want to compare (a) our \emph{mixed-strategy} GBR to 
(b) the standard GBR (where no message is ever sent directly
to the BS: they are always passed to the neighbour which has spent the
less energy amongst all lower neighbours (in terms of height).
We also compare the mixed-strategy to (c) the random strategy from \cite{nrj:balance:olivier} 
(see section: \ref{optimal solution})\\
\subsection{Simulation Results}
\begin{figure}[h]
\centering
  \includegraphics[angle=0,width=.4\textwidth]{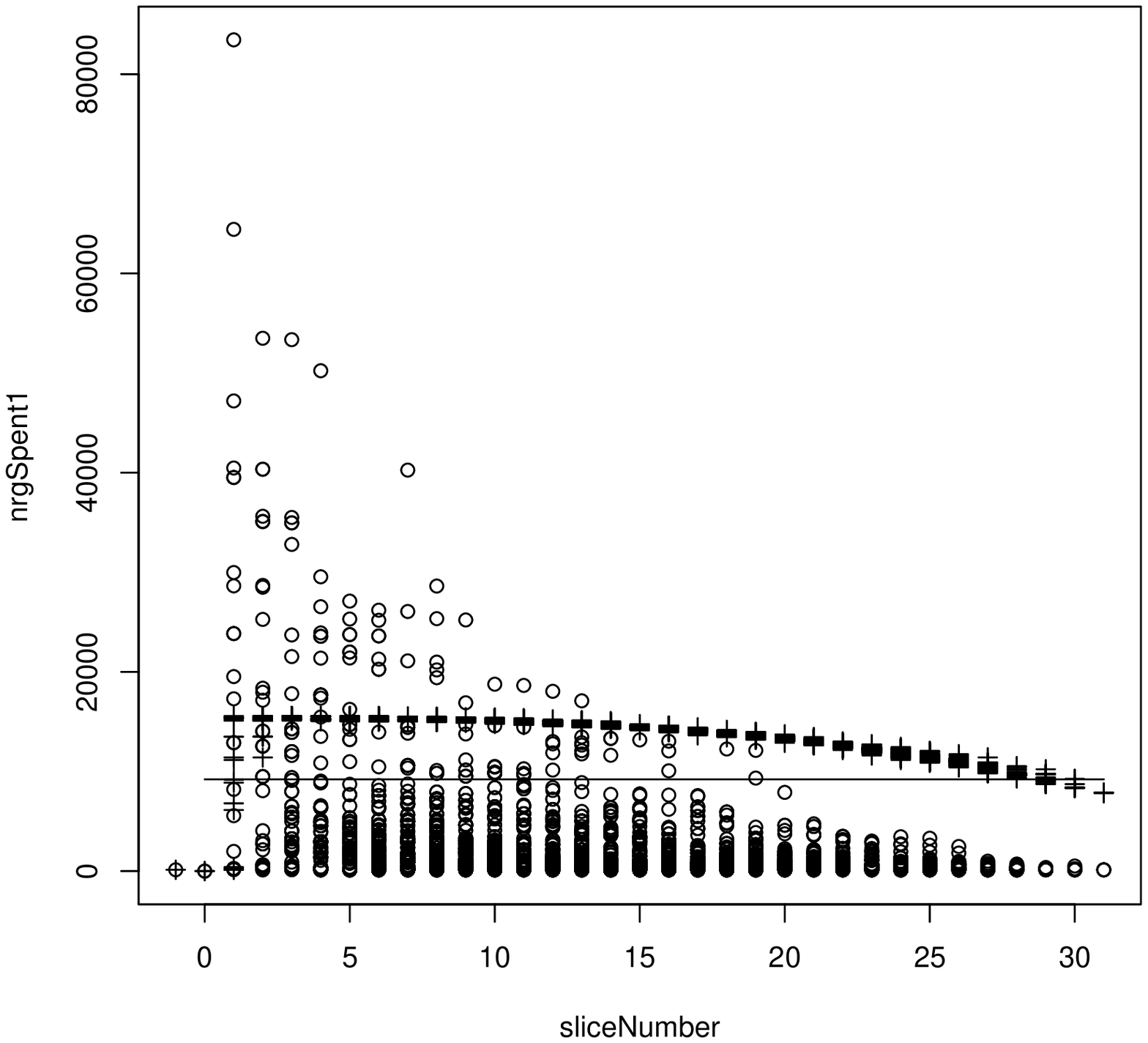}
\caption{}
\label{sliceVSnrg}
\end{figure}

\begin{figure}[h]
\centering
  \includegraphics[angle=0,width=.4\textwidth]{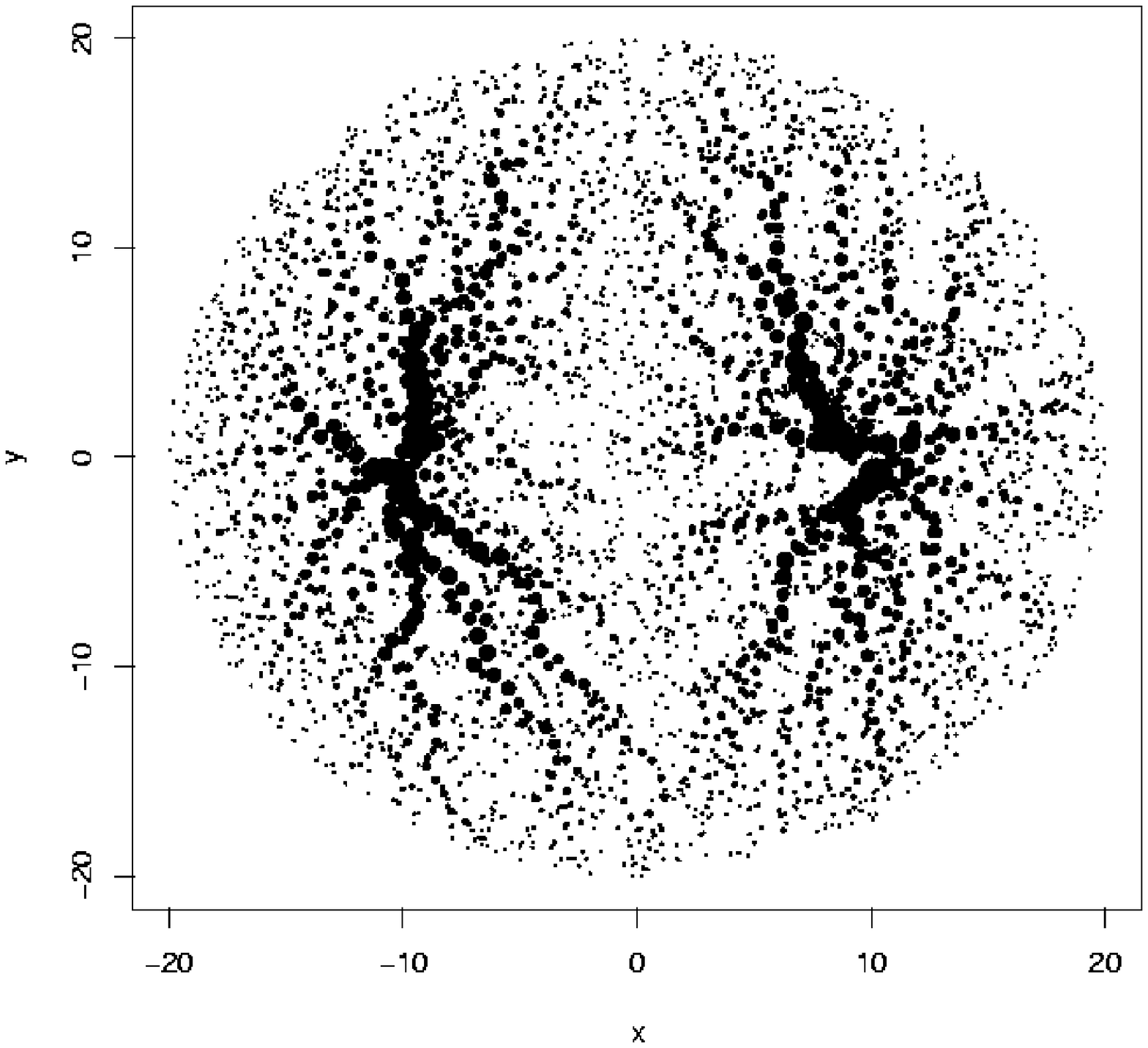}
\caption{}
\label{fractal}
\end{figure}

{\bf Scenario:} We consider a disc $D_R$ centered at the origin with
radius $R=20$, and we random-uniformly disperse $3769 = \lfloor 3 \Pi R^2\rfloor $ sensors in $D_R$.
Two base stations are placed at positions $(-10,0)$ and $(10,0)$. 
Next, we run the event-generation/message-propagation paradigm as described above for $5\cdot10^6$ rounds. 
The energy consumption for the three strategies is compared in figure \ref{sliceVSnrg}.
What we see on this figure is that the energy consumption is badly balanced for the standard GBR protocol:
a few nodes close to the sink spend a lot of energy, while the nodes away from the sink do not.
This can be further observed on figure \ref{fractal},
where a disk with surface proportional to the energy consumption (for the standard GBR protocol) 
is plotted for each sensor. 
This figure helps understand how GBR behaves: we see natural routes appearing which are
bottle-neck regions of the network, composed of central gateways through which most messages have to transit.
The traffic along these routes could probably be spread a bit better (e.g. by using a \emph{stream-based}
traffic spreading technique \cite{GBR}), 
but the main problem to energy balancing is that nodes close to the base stations use more energy than those away 
from the base stations, 
(as observed and explained in \cite{singh,nrj:balance:jose,nrj:balance:pierre,nrj:balance:olivier}).
On the contrary, figure \ref{sliceVSnrg} show that this problem is well addressed by the 
\emph{mixed-strategy} GBR protocol we propose.
\subsection{Comparison to an optimal solution}
\label{optimal solution}
In \cite{nrj:balance:olivier}, an offline algorithm producing an optimal\footnote{In terms of 
minimizing the maximum energy consumption amongst nodes} \emph{randomized-strategy} was proposed:
when a node
gets to transfer a message, it makes a random (according to a wisely chosen probability) 
decision on whether it should send it directly to the sink or pass it to a neighbour. 
In \cite{nrj:balance:olivier}, (like in \cite{singh,nrj:balance:jose,nrj:balance:pierre})
the analysis of the solution is statistical in the sense that for each slice only the average energy spent per sensor is
taken into account, and not the individual energies.
This is in fact equivalent to making the simplifying assumptions that
nodes at the same height share a common battery, and that messages can
be split in two smaller messages of arbitrary size (e.g. a message
could be split into a small sub-message
sent directly to the sink, while the rest of the \emph{same} message
is routed along the network).
Because of this implicitly simplifying statistic analysis,
(and also because it is computed offline) 
the randomized strategy should be considered as an ideal and unreachable optimal strategy. 
However, we can still compare our solution to it, 
and see on figure \ref{sliceVSnrg} that the \emph{mixed-strategy} GBR 
does not behave significantly worse than the \emph{randomized-strategy}.
\subsection{Statistics}
If we look at the maximum energy spent by a node for the standard GBR protocol, 
the mixed-strategy GBR protocol and the randomized-strategy we get values of $83448$, $15557$ and $9210$ respectively.
Observing figures \ref{fractal} and \ref{sliceVSnrg} and consistently with predictions
from \cite{singh,nrj:balance:jose,nrj:balance:pierre,nrj:balance:olivier}, wee see that
those nodes which spend the more energy are those
close to the sink.
These nodes seem to play the role of central gateways: when these
nodes run out of energy, the standard GBR routing is broken.
In light of this, it does not seem unreasonable to consider, as a measure of the lifespan
of the network, the worst-case energy consumption amongst nodes. 
For this measure we can numerically observe that in the above
experiment, the mixed-strategy GBR spends $1.7$ more worst-case energy
than the randomized-strategy, 
and that the standard GBR spends $5.4$ more worst-case energy than the
randomized strategy, and thus the mixed-strategy has a more
than $300\%$ lifespan increase when compared to the standard GBR.
Many other similar experiments at different scales (in terms of number
of nodes) and density give similar results.
\bibliography{bibSensor}
\end{document}